\documentclass[runningheads]{llncs}
\usepackage[utf8]{inputenc}
\usepackage[T1]{fontenc}
\usepackage[UKenglish]{babel}
\usepackage{amsmath}
\usepackage[dvipsnames]{xcolor}
\usepackage{amsfonts}
\usepackage{mathtools}
\usepackage{graphicx}
\usepackage{hyperref}
\usepackage[capitalize,noabbrev]{cleveref}
\usepackage{inconsolata}
\usepackage{bm}
\usepackage{todonotes}
\usepackage{listings}
\usepackage{microtype}
\usepackage{etoolbox}
\usepackage{verbatim}

\newcommand{\gram}{::=}
\newcommand{\Div}{\ |\ }
\newcommand{\kwd}[1]{\mathbf{#1}}
\newcommand{\kwdd}[1]{\bm{#1}}
\newcommand{\brak}[1]{\bm{\{}#1\bm{\}}}
\newcommand{\sqr}[1]{\bm{[\!\![\!\![}#1\bm{]\!\!]\!\!]}}
\newcommand{\rnd}[1]{\bm{(}#1\bm{)}}
\newcommand{\ang}[1]{\bm{\langle}#1\bm{\rangle}}
\newcommand{\ann}[1]{///@\textit{#1}}
\newcommand{\cfEn}[1]{{\color{Cerulean}\bm{[}\!\bm{[}}#1{\color{Cerulean}\bm{]}\!\bm{]}}}
\newcommand{\en}[1]{{\color{Fuchsia}\bm{\lceil}\!\!\bm{\lceil}}#1{\color{Fuchsia}\bm{\rceil}\!\!\bm{\rceil}}}
\newcommand{\opEn}[1]{{\color{RubineRed}\bm{(}\!\!\bm{(}}#1{\color{RubineRed}\bm{)}\!\!\bm{)}}}

\lstdefinelanguage{jolie}{
	tabsize = 1,
	keepspaces=true,
	mathescape=true,
	keywordstyle=\upshape\bfseries,
	keywords={type,interface,RequestResponse,undefined,int,void,string,range,length,enum,long,double,bool},
	morecomment=[f][\itshape][0]{///},
	literate = {\{}{{\upshape\bfseries\{}}1
	{\}}{{\upshape\bfseries\}}}1
	{\ )}{{\upshape\bfseries)}}1
	{(\ }{{\upshape\bfseries(}}1
	{?}{{\upshape\bfseries\hspace{0pt}?}}1
	{...}{{$\ldots$}}1
	}
\lstdefinelanguage{lemma}{
	tabsize = 1,
	mathescape=true,
	keepspaces=true,
	keywordstyle=\upshape\bfseries,
	keywords={context,structure,aggregate,applicationService,domainEvent,entity,factory,domainService,service,repository,specification,valueObject,infrastructureService,identifier,neverEmpty,part,procedure,function,closure,identifier,sideEffectFree,validator,collection,enum,int,long,double,string,unspecified,boolean},
	literate = {\{}{{\upshape\bfseries\{}}1
	{\}}{{\upshape\bfseries\}}}1
	{<}{{$\langle$}}1
	{>}{{$\rangle$}}1
	{\ )}{{\upshape\bfseries)}}1
	{(\ }{{\upshape\bfseries(}}1
	{...}{{$\ldots$}}1
}
\lstdefinestyle{xtend}{
	language=Java,
	tabsize=1,
	mathescape=true,
	keepspaces=true,
	morekeywords={val, override, def, it, typeof, map, join, dispatch, IF, ENDIF, empty, as},
	keywordstyle=\upshape\bfseries,	
	escapechar=\#,
	morestring=*[d]{'''},
	morestring=[s][]{«}{»},
	moredelim={[is][\itshape]{§§}{§§}},
	literate = {\{}{{\upshape\bfseries\{}}1
	{\}}{{\upshape\bfseries\}}}1
	{<}{{$\langle$}}1
	{>}{{$\rangle$}}1
	{\ )}{{\upshape\bfseries)}}1
	{(\ }{{\upshape\bfseries(}}1
	{...}{{$\ldots$}}1
}
\newcommand{\jolie}[1]{\lstinline[language=jolie]{#1}}
\newcommand{\Lemma}[1]{\lstinline[language=lemma]{#1}}

\usepackage{mdframed}
\usepackage{adjustbox}
\lstnewenvironment{listingLemma}
{%
	\lstset{language={lemma},basicstyle=\small}
	\mdframed[backgroundcolor=Cerulean!20,%
	innerleftmargin=1,%
	innerrightmargin=0,%
	innertopmargin=-3,%
	innerbottommargin=-10,%
	skipabove=.5ex, skipbelow=-.5ex%
	]
}
{\llap{\smash{\adjustbox{lap=.995\textwidth,raise=1.4em}{\color{gray}LEMMA}}}\endmdframed}
\lstnewenvironment{listingJolie}
{%
	\lstset{language={jolie},basicstyle=\small}
	\mdframed[backgroundcolor=Peach!20,%
	innerleftmargin=1,%
	innerrightmargin=0,%
	innertopmargin=-3,%
	innerbottommargin=-10,%
	skipabove=.5ex, skipbelow=.5ex%
	]
}
{\llap{\smash{\adjustbox{lap=.995\textwidth,raise=1.4em}{\color{gray}Jolie}}}\endmdframed}
\lstnewenvironment{listingXtend}[3]
{%
	\lstset{%
		style={xtend},%
		basicstyle=\small,%
		escapechar=\#,%
		frame=single,%
		float=t,%
		numbers=left,%
		tabsize=1,
		aboveskip=-0.1cm, 
		belowskip=-0.1cm, 
		morestring=[s][]{«}{»},
		morestring=*[d]{'''},
		showstringspaces=false,
		caption=#1,%
		label=#2,
		float=#3,
		literate=%
		{Ö}{{\"O}}1
		{Ä}{{\"A}}1
		{Ü}{{\"U}}1
		{ß}{{\ss}}1
		{ü}{{\"u}}1
		{ä}{{\"a}}1
		{ö}{{\"o}}1
		{~}{{\textasciitilde}}1
		{-}{-}1
		{«}{{\flqq}}1
		{»}{{\frqq}}1}%
}{}

\usepackage{tabularx}
\usepackage{array}
\newcolumntype{C}[1]{>{\centering\arraybackslash}p{#1}}
\newcolumntype{L}[1]{>{\raggedright\arraybackslash}p{#1}}
\usepackage{booktabs}

\usepackage{orcidlink}
\def\orcidID#1{\unskip$^{\orcidlink{#1}}$}

\def\toolname{LEMMA\-2\-Jolie}
\def\tool{\textsf{\toolname}}

\begin{document}
\title{Model-Driven Generation of Microservice Interfaces: From LEMMA Domain Models to Jolie APIs}
\titlerunning{Model-Driven Generation of Microservice Interfaces}

\author{
	Saverio Giallorenzo\inst{1}\orcidID{0000-0002-3658-6395}
	\and
	Fabrizio Montesi\inst{2}\orcidID{0000-0003-4666-901X}
	\and
	Marco Peressotti\inst{2}\orcidID{0000-0002-0243-0480}
	\and
	Florian Rademacher\inst{3}\orcidID{0000-0003-0784-9245}
}

\authorrunning{S.\ Giallorenzo, F.\ Montesi, M.\ Peressotti, and F.\ Rademacher}

\institute{
	Università di Bologna, Italy and INRIA, France
	\email{saverio.giallorenzo@gmail.com}
	\and
	University of Southern Denmark
	\email{\{fmontesi,peressotti\}@imada.sdu.dk}
	\and
	University of Applied Sciences and Arts Dortmund
	\email{florian.rademacher@fh-dortmund.de}
}

\hyphenation{Dev-Ops}

\maketitle

\begin{abstract}
	We formally define and implement a translation from domain models in the LEMMA modelling framework to microservice APIs in the Jolie programming language.
	Our tool enables a software development process whereby microservice architectures can first be designed with the leading method of Domain-Driven Design, and then corresponding data types and service interfaces (APIs) in Jolie are automatically generated. Developers can extend and use these APIs as guides in order to produce compliant implementations.
	Our tool thus contributes to enhancing productivity and improving the design adherence of microservices.
\end{abstract}

\section{Introduction}
Microservice Architecture (MSA) is one of the current leading patterns in distributed software architectures~\cite{Newman2015}.

While widely adopted, MSA comes with specific challenges regarding architecture design, development,
and operation~\cite{Dragoni2017,Soldani2018}. To cope with this complexity, researchers in software engineering and programming languages started proposing
linguistic approaches to MSA: language frameworks that ease the
design and development of MSAs with high-level constructs that make microservice concerns in the two different stages syntactically manifest.

Regarding development, Ballerina and Jolie are examples of programming languages
\cite{Oram2019,MGZ14} with new linguistic abstractions for effectively
programming the configuration and coordination of microservices. Regarding design, Model-Driven Engineering (MDE)~\cite{Combemale2017}
has gained relevance as a method for the specification of
service architectures~\cite{Ameller2015}, crystallised in MDE-for-MSA modelling languages such as MicroBuilder, MDSL, LEMMA, and
JHipster~\cite{Terzic2018,Kapferer2020b,Rademacher2020,JDL2022}.
Jolie's abstractions have been found to offer a productivity boost in industry~\cite{Guidi2019}. LEMMA provides linguistic support for the application of concepts from Domain-Driven Design~\cite{Evans2004,Rademacher2020}, and has been validated in real-world use cases~\cite{Sorgalla2021,Rademacher2021}.

Recently, it has been observed that the metamodels of LEMMA's modelling languages and the Jolie programming language have enough contact points to consider their integration~\cite{Giallorenzo2021}.
In the long term, such an integration could bring (quoting from~\cite{Giallorenzo2021})
\begin{quote}
	``\emph{an ecosystem that coherently combines MDE and programming abstractions to offer a tower of abstractions~\cite{M09} that supports a step-by-step refinement process from the abstract specification of a microservice architecture to its implementation}''. 
\end{quote}

The aim is to provide a toolchain that enables people to apply MDE to the design of microservices in LEMMA, and then seamlessly switch to a programming language with dedicated support for microservices like Jolie in order to develop an implementation of the design.
To this end, three important parts of the metamodels of LEMMA and Jolie need to be covered and integrated~\cite{Giallorenzo2021}:
\begin{enumerate}
	\item \emph{Application Programming Interfaces} (API), describing what functionalities (and their data types) a microservice offers to its clients;
	\item \emph{Access Points}, capturing where and how clients can interact with the API;
	\item \emph{Behaviours}, defining the internal business logic of a microservice.
\end{enumerate}

Since the API is the layer the other two build upon, in this paper we focus on concretising the relationship between LEMMA and Jolie API layers. To this end, we contribute a formal encoding between LEMMA's Domain Data Modelling Language (DDML) and Jolie types and interfaces. This encoding enables systematic translation of LEMMA domain models, which, following DDD principles, capture domain-specific types including operation signatures, to Jolie APIs. As a second contribution, we present \tool{}---a code generator that allows automatic translation of LEMMA domain models to Jolie APIs based on the introduced encoding. Specifically, \tool{} not only shows the encoding's feasibility and practicability, but also constitutes a crucial contribution towards improving the adoption of DDD in microservice design, which in practice is often perceived complex given the lack of formal guidelines on how to map DDD domain models to microservice code~\cite{Bogner2019}. 
We have evaluated \tool{} in the context of a nontrivial microservice architecture that had previously been used to validate LEMMA~\cite{Rademacher2021}, which covers all the aspects of the formal encoding. The generated Jolie code is as expected, in the sense that it is faithful to the formal encoding and the model defined in LEMMA.
We use snippets of this code to exemplify our method throughout the paper.

The remainder of the paper is organised as follows. \cref{sec:lemma-jolie-encoding} introduces and exemplifies the encoding between LEMMA's DDML and Jolie APIs. \cref{sec:tool} describes the architecture and implementation of \tool{}. \cref{sec:conclusion} presents future work and concludes the paper.

\section{Encoding LEMMA Domain Modelling Concepts in Jolie}\label{sec:lemma-jolie-encoding}
This section describes and exemplifies domain modelling with LEMMA (cf. \cref{sub:lemma-domain-concepts}), and the development of types and interfaces with Jolie (cf. \cref{sub:jolie-types-ifaces}). Next, it reports a formal encoding from LEMMA domain models to Jolie APIs and illustrates its application (cf. \cref{sec:encoding,sec:capturing-features}).

\subsection{LEMMA Domain Modelling Concepts}
\label{sub:lemma-domain-concepts}
LEMMA's DDML supports domain experts and service developers in the construction of models that capture domain-specific types of microservices. Figure~\ref{fig:lemmaGrammar} shows the core rules of the DDML grammar\footnote{The complete grammar can be found at \url{https://github.com/SeelabFhdo/lemma/blob/main/de.fhdo.lemma.data.datadsl/src/de/fhdo/lemma/data/DataDsl.xtext}.}.

\newcommand{\grayout}[1]{{\color{gray}{#1}}}
\begin{figure}[t]
	\[
	\begin{array}{lrl}
	CTX & \gram & \kwd{context}\ id\ \brak{ \overline{CT} }
	\\ CT & \gram & STR \Div COL \Div ENM
	\\ STR & \gram & \kwd{structure}\ id\ [\ang{\overline{STRF}}]\ \brak{\overline{FLD}\ \overline{OPS}}
	\\ STRF & \gram & \kwd{aggregate} \Div \kwd{domainEvent} \Div \kwd{entity} \Div \kwd{factory}
	\\      & \Div & \grayout{\kwd{service}} \Div \grayout{\kwd{repository}} \Div \kwd{specification} \Div \kwd{valueObject}
	\\ FLD & \gram & id\ id\ [\ang{\overline{FLDF}}] \Div S\ id\ [\ang{\overline{FLDF}}]
	\\ FLDF & \gram & \kwd{identifier} \Div \kwd{part}
	\\ OPS & \gram & \kwd{procedure}\ id\ [\ang{\overline{OPSF}}]\ \rnd{\overline{FLD}} 
	\Div \kwd{function}\ (id \Div S)\ id\ [\ang{\overline{OPSF}}]\ \rnd{\overline{FLD}}
	\\ OPSF & \gram & \grayout{\kwd{closure}} \Div \kwd{identifier} \Div \grayout{\kwd{sideEffectFree}} \Div \kwd{validator}
	\\ COL & \gram & \kwd{collection}\ id\ \brak{(S \Div id)}
	\\ ENM & \gram & \kwd{enum}\ id\ \brak{\overline{id}}
	\\ S & \gram & \kwd{int} \Div \kwd{string} \Div \kwd{unspecified} \Div \dots
	\end{array}
	\]
	\caption{Simplified grammar of LEMMA's DDML. Greyed out features are out of the scope of this paper and subject to future work.}
	\label{fig:lemmaGrammar}
\end{figure}

The DDML follows DDD to capture domain concepts. DDD's Bounded Context pattern~\cite{Evans2004} is crucial in MSA design as it makes the boundaries of coherent domain concepts explicit, thereby defining their scope and applicability~\cite{Newman2015}. A LEMMA domain model defines named bounded \(\kwd{contexts}\) (rule~\(CTX\) in \cref{fig:lemmaGrammar}). A \(\kwd{context}\) may specify domain concepts in the form of complex types (\(CT\)), which are either structures (\(STR\)), collections (\(COL\)), or enumerations (\(ENM\)).

A \(\kwd{structure}\) gathers a set of data fields (\(FLD\)). The type of a data field is either a complex type from the same bounded context (\(id\)) or a built-in primitive type, e.g., \(\kwd{int}\) or \(\kwd{string}\) (\(S\)). The \(\kwd{unspecified}\) keyword enables continuous domain exploration according to DDD~\cite{Evans2004}. That is, it supports the construction of underspecified models and their subsequent refinement as one gains new domain knowledge~\cite{Rademacher2020c}. Next to fields, \(\kwd{structures}\) can comprise operation signatures (\(OPS\)) to reify domain-specific behaviour. An operation is either a \(\kwd{procedure}\) without a return type, or a \(\kwd{function}\) with a complex or primitive return type.

LEMMA's DDML supports the assignment of DDD patterns, called \emph{features}, to
structured domain concepts and their components. For instance, the
\(\kwd{entity}\) feature (rule~\(STRF\) in \cref{fig:lemmaGrammar})
expresses that a structure comprises a notion of domain-specific identity. The
\(\kwd{identifier}\) feature then marks the data fields (\(FLDF\))
or operations (\(OPSF\)) of an \(\kwd{entity}\) which
determine its identity. For compactness, we defer the detailed
presentation of the considered DDD features to \cref{sec:capturing-features}, when
discussing their relationship with our encoding to Jolie.

The DDML also enables the modelling of \(\kwd{collection}\)s (rule~\(COL\) in
\cref{fig:lemmaGrammar}), which represent sequences of primitives (\(S\)) or
complex (\(id\)) values, as well as \(\kwd{enum}\)erations (\(ENM\)), which gather sets of predefined literals.

The following listing shows an example of a LEMMA domain model constructed with the grammar of the DDML~\cite{Rademacher2021}.

\begin{listingLemma}
	context BookingManagement {
		structure ParkingSpaceBooking<entity> {
			long bookingID<identifier>,
			double priceInEuro,
			function double priceInDollars
		}
	}
\end{listingLemma}

The domain model defines the bounded \(\kwd{context}\) {\itshape\rmfamily Book\-ing\-Man\-age\-ment} and its \(\kwd{structure}\)d domain concept {\itshape\rmfamily Park\-ing\-Space\-Book\-ing}. It is a DDD \(\kwd{entity}\) whose {\itshape\rmfamily book\-ing\-ID} field holds the \(\kwd{identifier}\) of an entity instance. The entity also clusters the field {\itshape\rmfamily price\-In\-Euro} to store the price of a parking space booking, and the \(\kwd{function}\) signature {\itshape\rmfamily price\-In\-Dol\-lars} for currency conversion of a booking's price.

\subsection{Jolie Types and Interfaces}
\label{sub:jolie-types-ifaces}

Jolie interfaces and types define the functionalities of a microservice and the data types associated with those functionalities i.e., the API of a microservice.
\cref{fig:jolieGrammar} shows a simplified variant of the grammar of
Jolie APIs, taken from~\cite{MGZ14} and updated to Jolie 1.10 (the
latest major release at the time of writing).

\begin{figure}[htbp]
	\[
	\begin{array}{lll}
	I & \gram & \kwd{interface}\ id\ \brak{ \overline{ \kwd{RequestResponse}\ id\rnd{TP_1}\rnd{TP_2} }}
	\\ TP & \gram & id \Div B
	\\ TD & \gram & \kwd{type}\ id:\ T
	\\ T & \gram & B\ [ \brak{ \overline{id\ C:\ T} } ] \Div \kwd{undefined}
	\\ C & \gram & \sqr{min, max} \Div \kwdd{*} \Div \kwdd{?}
	\\ B & \gram & \kwd{int}[\rnd R] \Div \kwd{string}[\rnd R] \Div \kwd{void} \Div \dots
	\\ R & \gram & \kwd{range}\rnd{\sqr{min,max}} \Div \kwd{length}\rnd{\sqr{min,max}} \Div \kwd{enum}\rnd{...} \Div \dots
	\end{array}
	\]
	\caption{Simplified syntax of Jolie APIs (types and interfaces)}
	\label{fig:jolieGrammar}
\end{figure}

An \(\kwd{interface}\) is a collection of named operations (\(\kwd{RequestResponse}\)), where the sender delivers its message of type \(TP_1\) and waits for the receiver to reply with a response of type \(TP_2\)---although Jolie also supports \(\kwd{oneWay}\)s, where the sender delivers its message
to the receiver, without waiting for the latter to process it (fire-and-forget), we omit them here because they are not used in the encoding (cf. \cref{sec:encoding}).
Operations have types describing the shape of the data structures they can exchange, which can either define custom, named types (\(id\)) or basic ones (\(B\))
(\(\kwd{int}\)egers, \(\kwd{string}\)s, etc.).

Jolie \(\kwd{type}\) definitions (\(TD\)) have a tree-shaped structure. At their root, we
find a basic type (\(B\))---which can include a refinement (\(R\)) to express
constraints that further restrict the possible inhabitants of the type~\cite{FP91}.
The possible branches of a \(\kwd{type}\) are a set of nodes, where each node
associates a name (\(id\)) with an array with a range length (\(C\)) and a type
\(T\).

Jolie data types and interfaces are technology agnostic: they model Data
Transfer Objects (DTOs) built on native types generally available in most
architectures~\cite{Daigneau2012}.

Based on the grammar in \cref{fig:jolieGrammar}, the following listing shows the Jolie equivalent of the example LEMMA domain model from \cref{sub:lemma-domain-concepts}.

\begin{listingJolie}
	///@beginCtx(BookingManagement)
	///@entity
	type ParkingSpaceBooking {
		///@identifier
		bookingID: long
		priceInEuro: double
	}
	interface ParkingSpaceBooking_interface {
		RequestResponse:
			priceInDollars(ParkingSpaceBooking)(double)
	}
	///@endCtx
\end{listingJolie}

Structured LEMMA domain concepts like {\itshape\rmfamily Park\-ing\-Space\-Book\-ing} and their data fields, e.g., {\itshape\rmfamily book\-ing\-ID}, are directly translatable to corresponding Jolie \(\kwd{type}\)s.

To map LEMMA DDD information to Jolie, we use Jolie documentation comments ({\itshape\rmfamily ///}) together with an {\itshape\rmfamily @}-sign. It is followed by (i) the string {\itshape\rmfamily be\-gin\-Ctx} and the parenthesised name of a modelled bounded context, e.g., {\itshape\rmfamily Book\-ing\-Man\-age\-ment}; (ii) the DDD feature name, e.g., {\itshape\rmfamily entity}; or (iii) the string {\itshape\rmfamily end\-Ctx} to conclude a bounded context. This approach enables to preserve semantic DDD information for which Jolie currently does not support native language constructs. The comments serve as documentation to the programmer who will implement the API. In the future, we plan on leveraging these special comments also in automatic tools (see \cref{sec:capturing-features,sec:conclusion}).

LEMMA operation signatures are expressible as \(\kwd{Re\-quest\-Re\-sponse}\) operations within a Jolie \(\kwd{interface}\) for the LEMMA domain concept that defines the signatures. For example, we mapped the domain concept {\itshape\rmfamily Park\-ing\-Space\-Book\-ing} and its operation signature {\itshape\rmfamily price\-In\-Dol\-lars} to the Jolie interface {\itshape\rmfamily ParkingSpaceBooking\_interface} with the operation {\itshape\rmfamily price\-In\-Dol\-lars}.

\subsection{Encoding LEMMA Domain Models as Jolie APIs}
\label{sec:encoding}

In the following, we report an encoding from LEMMA domain models to Jolie APIs that formalises and extends the mapping exemplified in \cref{sub:jolie-types-ifaces}. \cref{fig:encoding} shows the encoding.

The encoding is split in three encoders: the \emph{main} encoder
\(\cfEn{\cdot}\) walks through the structure of LEMMA domain models to generate
Jolie APIs using the encoders for \emph{operations} (\(\opEn{\cdot}\)) and for
\emph{structures} (\(\en{\cdot}\)), respectively.

The operations encoder \(\opEn{\cdot}\) generates Jolie interfaces based on
\(\kwd{procedure}\)s and \(\kwd{function}\)s in the given models by translating
structure-specific operations into Jolie operations. 
This translation requires some care. On one hand, LEMMA's
\(\kwd{procedure}\)s and \(\kwd{function}\)s are similar in nature to methods of
OOP, since they operate on data stored in their defining structure. On the other
hand, Jolie does not support objects in the OOP sense but rather separates data
from code that can operate on it (operations).
Therefore, the encoding needs to decouple \(\kwd{procedure}\)s
and \(\kwd{function}\)s from their defining structures
as illustrated in \cref{sub:jolie-types-ifaces} by the mapping of the LEMMA
domain concept {\itshape\rmfamily Park\-ing\-Space\-Book\-ing} and its operation signature
{\itshape\rmfamily price\-In\-Dol\-lars} to the Jolie interface
{\itshape\rmfamily ParkingSpaceBooking\_interface} with the operation {\itshape\rmfamily price\-In\-Dol\-lars}.

Given a structure \(X\), we extend the signature of its \(\kwd{procedure}\)s
with a parameter for representing the structure they act on and a return type
\(X\) for the new state of the structure, essentially turning them into
functions that transform the enclosing structure. For instance, we regard a
procedure with signature \((Y \times \dots \times Z)\) in \(X\) as a function
with type \(X \times Y \times \dots \times Z \rightarrow X\). This approach is
not new and can be found also in modern languages like Rust~\cite{KN19,rust-language-ref} and
Python~\cite{python-language-ref}. The operation synthesised by the \(\opEn{\cdot}\)
encoder accepts the \(id\_type\) generated by the \(\cfEn{\cdot}\) encoder that,
in turn, has a \(\mathit{self}\) leaf carrying the enclosing data structure
(\(id_s\)). The encoding of \(\kwd{function}\)s follows a similar path. Note
that, when encoding \(\mathit{self}\) leaves, we do not impose the constraint of
providing one such instance (represented by the \jolie{?} cardinality), but
rather allow clients to provide it (and leave the check of its presence to the
API implementer).

The main encoder \(\cfEn{\cdot}\) and the structure encoder \(\en{\cdot}\)
transform LEMMA types into Jolie types. \Lemma{context}s translate into
pairs of \(\ann{beginCtx}(context\_name)\) and \(\ann{endCtx}\) Joliedoc comment annotations. All the other constructs translate into \jolie{type}s and their subparts. When
translating \Lemma{procedure}s and \Lemma{function}s, the two encoders follow
the complementary scheme of \(\opEn{\cdot}\) and synthesise the types for the
generated operations. The other rules are straightforward.

\begin{figure}[t]
	\newcommand{\fSkip}{.7em}
	\resizebox*{\textwidth}{!}{$
		\begin{array}{lll}
		\cfEn{\kwd{context}\ id\ \brak{\overline{CT}}} & = & \ann{beginCtx}\rnd{id}\\
		&& \overline{\cfEn{CT}}\\
		&&\ann{endCtx}
		\\[\fSkip] \opEn{\kwd{structure}\ id\ [\ang{\overline{STRF}}]\ \brak{\overline{FLD}\ \overline{OPS}}} & = & [\overline{\ann{STRF}}]\ \kwd{interface}\ id\_\mathit{interface}\ \brak{\overline{\opEn{OPS}_{id}}}
		\\[\fSkip] \opEn{\kwd{procedure}\ id\ [\ang{\overline{OPSF}}]\ \rnd{\overline{FLD}}}_{id_s} & = & \kwd{RequestResponse}:\ 
		[\overline{\ann{OPSF}}]\ 
		id\rnd{id\_type}\rnd{id_s}\
		\\[\fSkip] \opEn{\kwd{function}\ ( S \Div id_r)\ id\ [\ang{\overline{OPSF}}]\ \rnd{\overline{FLD}}}_{id_s} & = & \kwd{RequestResponse}:\ 
		[\overline{\ann{OPSF}}]\ 
		id\rnd{ id\_type }\rnd{(\en{S} \Div id_r)}
		\\[\fSkip] \cfEn{\kwd{structure}\ id\ [\ang{\overline{STRF}}]\ \brak{\overline{FLD}\ \overline{OPS}}} & = & 
		\begin{array}[t]{l}
		\kwd{type}\ \en{\kwd{structure}\ id\ [\ang{\overline{STRF}}]\ \brak{\overline{FLD}}}\
		\\ \overline{\cfEn{OPS}}_{id}\ \opEn{\kwd{structure}\ id\ [\ang{\overline{STRF}}]\ \brak{\overline{OPS}}}_{id}
		\end{array}
		\\[2em] \cfEn{\kwd{procedure}\ id\ [\ang{\overline{OPSF}}]\ \rnd{\overline{FLD}}}_{id_s} & = & \kwd{type}\ id\_type:\ \kwd{void}\ \brak{ \mathit{self}?:\ id_s\ \overline{\en{FLD}}}\
		\\[\fSkip] \cfEn{\kwd{function}\ (id_r \Div S) \ id\ [\ang{\overline{OPSF}}]\ \rnd{\overline{FLD}}}_{id_s} & = & \kwd{type}\ id\_type:\ \kwd{void}\ \brak{\mathit{self}?:\ id_s\ \overline{\en{FLD}}}
		\\[\fSkip] \cfEn{\kwd{collection}\ id\ \brak{(S \Div id_r )}} & = & \kwd{type}\ id:\ \kwd{void}\ \brak{ \en{\kwd{collection}\ id\ \brak{(S \Div id_r)}} }
		\\[\fSkip] \cfEn{\kwd{enum}\ id\ \brak{\overline{id}}} & = & \kwd{type}\ \en{\kwd{enum}\ id\ \brak{\overline{id}}}
		\\[\fSkip] \en{\kwd{structure}\ id\ [\ang{\overline{STRF}}]\ \brak{\overline{FLD}}} & = & [\overline{\ann{STRF}}]\ id:\ \kwd{void}\ \brak{\overline{\en{FLD}}} 
		\\[\fSkip] \en{S\ id\ [\ang{\overline{FLDF}}]} & = & [\overline{\ann{FLDF}}]\ id:\ \en{S}
		\\[\fSkip] \en{id_r\ id\ [\ang{\overline{FLDF}}]} & = & [\overline{\ann{FLDF}}]\ id:\ id_r
		\\[\fSkip] \en{\kwd{collection}\ id\ \brak{S}} & = & id\kwd{*}: \ \en{S} 
		\\[\fSkip] \en{\kwd{collection}\ id\ \brak{id_r}} & = & id\kwd{*}: \ id_r
		\\[\fSkip] \en{\kwd{enum}\ id\ \brak{\overline{id}}} & = & id:\ \kwd{string}\rnd{\mathit{enum}(\overline{``id''})}
		\\[\fSkip] \en{\kwd{int}} & = & \kwd{int}
		\\[\fSkip] \en{\kwd{unspecified}} & = & \kwd{undefined}
		\end{array}
		$}
	\caption{Salient parts of the Jolie encoding for LEMMA's domain modelling concepts.}
	\label{fig:encoding}
\end{figure}

\subsection{Applying the Encoding}
\label{sec:capturing-features}

This subsection illustrates the application of the encoding from \cref{sec:encoding} using the Booking Management Microservice (BMM) of a microservice-based Park and Charge Platform (PACP) modelled with LEMMA~\cite{Rademacher2021}. The PACP enables drivers of electric vehicles to offer their charging stations for use by others. Its BMM manages the corresponding bookings based on domain concepts that were designed following DDD principles~\cite{Evans2004} and expressed in LEMMA's DDML.

In the following paragraphs, unless indicated, the encoded Jolie APIs respect the DDD constraints expressed by the considered features.

\subsubsection{Aggregate and Part}
In DDD, aggregates prescribe object graphs, whose parts must maintain a consistent state \cite{Evans2004}. Aggregates are always loaded from and stored to a database in a consistent state and within one transaction. A DDD aggregate consists of at least an entity or value object (see below). The following left listing shows the {\itshape\rmfamily PSB} aggregate in the LEMMA domain model for the BMM.

\noindent\begin{minipage}{.5\textwidth}
\begin{listingLemma}
structure PSB 
< aggregate > {
	TimeSlot timeSlot < part >,
	double priceInEuro
}
structure TimeSlot { ... }
$\vphantom{ciao}$
\end{listingLemma}
\end{minipage}
\begin{minipage}{.5\textwidth}
\begin{listingJolie}
///@aggregate
type PSB {
	///@part
	timeSlot: TimeSlot
	priceInEuro: double
}
type TimeSlot { ... }
\end{listingJolie}
\end{minipage}

\Lemma{PSB} is a \Lemma{structure}d domain concept with the \(\kwd{aggregate}\) feature (cf. \cref{sub:lemma-domain-concepts}) and it clusters the field {\itshape\rmfamily time\-Slot}, which has a structured type and is a \(\kwd{part}\) of the aggregate. Notice that for this domain model, LEMMA's DDML would emit warnings, because (i) a DDD aggregate must specify a root entity; and (ii) a part should either be an entity or value object~\cite{Evans2004}. We extend the {\itshape\rmfamily PSB} aggregate below to gradually fix these issues, thereby explaining the semantics of DDD entities and value objects.

In the Jolie encoding (on the right), we have as many \jolie{type} definitions as we have \Lemma{structure}s in the LEMMA model. 

\subsubsection{Entity and Identifier}
\label{sec:entity}

Instances of DDD entities are distinguishable by a domain-specific identity~\cite{Evans2004}, e.g., a unique ID. The following left listing extends the {\itshape\rmfamily PSB} aggregate with the \(\kwd{entity}\) feature and an \(\kwd{identifier}\) field.

\noindent\begin{minipage}{.5\textwidth}
\begin{listingLemma}
structure PSB 
< aggregate, entity > {
	long bookingID < identifier >,
	TimeSlot timeSlot < part >,
	double priceInEuro
}
$\vphantom{ciao}$
$\vphantom{ciao}$
$\vphantom{ciao}$
\end{listingLemma}
\end{minipage}
\begin{minipage}{.5\textwidth}
\begin{listingJolie}
///@aggregate
///@entity
type PSB {
	///@identifier
	bookingID: long
	///@part
	timeSlot: TimeSlot
	priceInEuro: double
}
\end{listingJolie}
\end{minipage}

LEMMA's DDML requires the \(\kwd{entity}\) feature on an aggregate to signal that its fields prescribe the structure of its root entity. The \(\kwd{identifier}\) feature can be used to mark those fields that determine the identity of an entity instances. In the example above, the value of {\itshape\rmfamily book\-ing\-ID} is marked to identify {\itshape\rmfamily PSB}s.

The Jolie encoding of \Lemma{entity} and \Lemma{identifier} fields is straightforward.

\looseness=-1
Next to fields, DDML supports the \(\kwd{identifier}\) feature on a
single \(\kwd{function}\) of an entity to enable identity calculation at runtime.
To illustrate this approach, the following listing models the {\itshape\rmfamily
	book\-ing\-ID} of the {\itshape\rmfamily PSB} root
entity as a function.

\noindent\begin{minipage}{.5\textwidth}
\begin{listingLemma}
structure PSB < entity > {
	function long bookingID 
	< identifier > (  ),
	...
}
$\vphantom{ciao}$
$\vphantom{ciao}$
$\vphantom{ciao}$
\end{listingLemma}
\end{minipage}
\noindent\begin{minipage}{.5\textwidth}
\begin{listingJolie}
///@entity
type PSB  {	... }
type bookingID_type { self?: PSB }
interface PSB_interface {
	RequestResponse:
		///@identifier
		bookingID( bookingID_type )( long )
}
\end{listingJolie}
\end{minipage}

Following our encoding (cf. \cref{sec:encoding}), we create the Jolie \jolie{type}
\jolie{bookingID_type} for the {\itshape\rmfamily book\-ing\-ID} \Lemma{identifier}
\Lemma{function}. The \jolie{type}'s \jolie{self} leaf enables implementers
to access the fields of the \jolie{PSB} and define how to compute the identifier.

\subsubsection{Factory}
DDD factories make the creation of objects with complex consistency requirements explicit~\cite{Evans2004}. LEMMA's DDML considers factories to constitute \(\kwd{function}\)s that return instances of aggregates, entities, or value objects. The following left listing illustrates the usage of factories by specifying the \(\kwd{factory}\) function {\itshape\rmfamily cre\-ate} as part of the {\itshape\rmfamily PSB} aggregate. This function shall create {\itshape\rmfamily PSB} instances for a given time slot {\itshape\rmfamily timeSlot} and a {\itshape\rmfamily priceInEuro}.

\noindent\begin{minipage}{.5\textwidth}
\begin{listingLemma}
structure PSB < ... >{
	TimeSlot timeSlot,
	double priceInEuro,
	function PSB create<factory>(
		TimeSlot timeSlot, 
		double priceInEuro
	)
	$\vphantom{ciao}$
	$\vphantom{ciao}$
}
\end{listingLemma}
\end{minipage}
\begin{minipage}{.5\textwidth}
\begin{listingJolie}
type PSB { ... }
///@factory
type create_type {
	timeSlot: TimeSlot
	priceInEuro: double
}
interface PSBFactory_interface {
	RequestResponse:
		create(create_type)(PSB)
}
\end{listingJolie}
\end{minipage}

As opposed to the encoding for LEMMA \Lemma{identifier} \Lemma{function}s (see above),
we do not encode a \jolie{self} leaf in Jolie \jolie{type}s such as \jolie{create\_type}
for LEMMA \(\kwd{factory}\) \Lemma{function}s. Since the semantics of factories is that of generating an instance of
the enclosing \Lemma{structure}, it would not make sense to pass to it one of
those instances as a \jolie{self} leaf. Consequently, we
could include a rule in \cref{fig:encoding} which avoids the generation of said
\jolie{self} leaf (this is more an issue of minimality of the generated code,
since we set the leaf as optional (\jolie{?})). Additionally, we can enforce a
check on Jolie operations like \jolie{create} following immediately after \(\ann{factory}\)-commented \jolie{type}s
by making sure their input \jolie{type}s do not contain the produced \jolie{type}, e.g., \jolie{PSB}.
Complementary, we can also check that the response type of Jolie-encoded factory operations
coincides with the produced \jolie{type}.

\subsubsection{Specification and Validator}
DDD specifications are domain concepts that make business rules, policies, or consistency specifications for aggregates explicit~\cite{Evans2004}. A specification must comprise one or more validators, which are functions with a boolean return type that reify the specification's predicates.

LEMMA's DDML provides the features \(\kwd{specification}\) and \(\kwd{validator}\) to mark structures as specifications and identify their validators. The following left listing extends the BMM's domain model with the {\itshape\rmfamily Book\-ing\-Ex\-pir\-a\-tion} specification. Its {\itshape\rmfamily is\-Ex\-pired} validator returns \(\kwd{true}\) if a parking space booking in the form a {\itshape\rmfamily PSB} instance has expired.

\noindent\begin{minipage}{.5\textwidth}	
\begin{listingLemma}
structure PSB < ... > { ... }
structure BookingExpiration 
< specification > {
	function boolean isExpired
	< validator > ( PSB p )
}
$\vphantom{ciao}$
$\vphantom{ciao}$
\end{listingLemma}%
\end{minipage}
\begin{minipage}{.5\textwidth}
\begin{listingJolie}	
type PSB { ... }
///@specification
type isExpired_type { p: PSB }
interface BookingExpiration_interface {
	RequestResponse:
		///@validator
		isExpired(isExpired_type)(bool)
}
\end{listingJolie}
\end{minipage}

Since the specification is a field-less \Lemma{structure}, we do not create a
corresponding \jolie{type} \jolie{BookingExpiration} as it would be empty. Instead,
and as per our encoding (cf. \cref{sec:encoding}), we create the
\(\ann{specification}\)-annotated \jolie{type} \jolie{isExpired_type} for the
{\itshape\rmfamily is\-Ex\-pired} \Lemma{validator} within the \jolie{interface}
\jolie{BookingExpiration_interface}.
From the point of view of the consistency of the annotations, following the namespace
convention from \cref{fig:encoding}, we can check that the \(\ann{validator}\)
actually accepts the related \Lemma{structure}. To do this, we follow the
``breadcrumbs'' left by our encoders. First, we find a
\(\ann{validator}\)-commented \jolie{RequestResponse} (e.g.,
\jolie{isExpired}) and we make sure its response type is \jolie{bool}.
Then, we follow the request type (e.g., \jolie{isExpired_type}) to make
sure that: \emph{i}) the \(\ann{validator}\) has an associated
\(\ann{specification}\) (e.g., \jolie{isExpired_type}) \jolie{type} and
\emph{ii}) the \jolie{type} has one leaf, which
is the \Lemma{structure} the \Lemma{validator} validates. 

Notice that, here, we lose the enclosing relation between
\jolie{BookingExpiration} and \jolie{PSB}. This might introduce subtle bugs
(e.g., due to typos), since we have only one place (the request type of the
\(\ann{validator}\)) that states the relation between the \Lemma{specification}
and the \Lemma{validator}. To strengthen our checks, we might include a rule in
\cref{fig:encoding} which would include the reference to \jolie{PSB} in the
annotation comment of the \Lemma{validator}, e.g.,
\(\ann{validator}\)\jolie{(PSB)}. In this way, we can assert the
correspondence between the validated \jolie{type} and the type of the
other leaf in the request of the \(\ann{validator}\).

\subsubsection{Value Object and Domain Event}
\label{sec:valueObject}
As opposed to entities, DDD value objects cluster data and logic, which are not dependent on objects' identity~\cite{Evans2004}.
Thus, value objects serve as DTOs for data exchange between microservices~\cite{Newman2015}. In asynchronous communication scenarios, value objects can model domain events emitted by a bounded context during runtime~\cite{Evans2004}. For example, all PACP microservices interact with each other via domain events~\cite{Rademacher2021}.

LEMMA's DDML supports the \(\kwd{valueObject}\) and \(\kwd{domainEvent}\) features to mark structured domain concepts as value objects and possibly as domain events. The following left listing illustrates the usage of the \(\kwd{valueObject}\) feature.

\noindent\begin{minipage}{.5\textwidth}
\begin{listingLemma}
context BookingManagement {
	structure PSB < ... > {
		TimeSlot timeSlot,
		double priceInEuro
	}
	structure PSB_VO< valueObject > {
		TimeSlot timeSlot,
		double price,
		string currency
	}
	structure TimeSlot< valueObject > { 
		... 
	}
}
\end{listingLemma}
\end{minipage}
\begin{minipage}{.5\textwidth}
\begin{listingJolie}
///@beginCtx(BookingManagement)
type PSB { 
	timeSlot: TimeSlot
	priceInEuro: double
}
///@valueObject
type PSB_VO {
	timeSlot: TimeSlot
	price: double
	currency: string
}
///@valueObject
type TimeSlot { ... }
///@endCtx
\end{listingJolie}
\end{minipage}

Above, we extend the BMM's domain model with the {\itshape\rmfamily PSB\_VO} value object: a DTO for the {\itshape\rmfamily PSB} aggregate that slightly changes it type to make its representation more general. Namely, {\itshape\rmfamily PSB\_VO} makes the {\itshape\rmfamily currency} explicit and separates it from the value of the {\itshape\rmfamily priceInEuro}, which we store in the field {\itshape\rmfamily price}. The {\itshape\rmfamily time\-Slot} field remains the same, but we make sure it is also a \Lemma{valueObject}.

The LEMMA domain model also shows the definition of bounded \(\kwd{context}\)s in the DDML. All three structures {\itshape\rmfamily PSB}, {\itshape\rmfamily PSB\_VO}, and {\itshape\rmfamily Time\-Slot} are enclosed by the {\itshape\rmfamily Book\-ing\-Man\-age\-ment} \(\kwd{context}\) on which the BMM operates exclusively.

The encoding from LEMMA to Jolie follows \cref{fig:encoding} without exceptions.
Notice, in particular, the ``opening'' \(\ann{beginCtx}\mathit{(BookingManagement)}\)
and ``closing'' \(\ann{endCtx}\) comments for the context. With those
comments, we are declaring that the types (and interfaces) that appear
between them belong to the context \jolie{BookingManagement}. In LEMMA, contexts
indicate a boundary within which (complex) types belonging in the same context
can co-exist and interact (e.g., by being part of the inputs and output of
\Lemma{procedure}s and \Lemma{function}s). Then, as seen above,
\Lemma{valueObject}s exist to allow data to cross boundaries, by defining
data types (e.g., \Lemma{structure}s) purposed to act as DTOs.

While the encoding from LEMMA's DDML ensures that, at the API level, the
anti-corruption invariants defined by \Lemma{context}s and \Lemma{valueObject}s
are preserved (e.g., there exists no \jolie{type} with leaves whose types belong
in different contexts nor \jolie{interface}s belonging in a context that accept
types from another context, unless \(\ann{valueObject}\)s), this is not the case
for behaviour, which can arbitrarily combine data structures and operators.

A possible way to ensure the enforcement of LEMMA's DDML anti-corruption
invariants also in behaviours (which will be subject to future work), is through
the definition of static checks that trace the contexts in which values
belong---from the types of the operations that generated them, via
receptions---and prohibit mixing values that belong in different contexts (e.g.,
by forbidding to use them with operations belonging in different contexts,
although their types might be compatible). This static check would also handle
the exception of values whose types are annotated as \(\ann{valueObject}\)s,
which are the only ones allowed to be used in a mixed way (i.e., in operations
that take or produce \(\ann{valueObject}\)-annotated types.).

\subsubsection{Additional Features}

Our encoding captures the \Lemma{repository} and \Lemma{closure} features of
LEMMA's DDML (cf. \cref{fig:lemmaGrammar}) without exceptions. Checks regarding
the {\rmfamily\bfseries side\-Ef\-fect\-Free} feature follow the same
considerations of the \Lemma{valueObject} feature: we need to inspect a service
behaviour to make sure its does not modify the values obtained from
\(\ann{sideEffectFree}\)-commented operations. \Lemma{service}s are a
generalisation of the \Lemma{specification} feature, where we have a structure
that contains only \Lemma{function}s and \Lemma{procedure}s---LEMMA further
refines \Lemma{service}s into, \Lemma{domainService}s,
\Lemma{infrastructureService}s, \Lemma{applicationService}s, which are subject
to future works.

\section{\toolname: A Code Generator to Derive Jolie APIs from LEMMA Domain Models}\label{sec:tool}

This section presents our \tool{} tool which makes the encoding presented in \cref{sec:lemma-jolie-encoding} practically applicable. In the sense of MDE, \tool{} is a \emph{model-to-text transformation}~\cite{Combemale2017} that generates Jolie APIs from LEMMA domain models. \cref{sub:tool-architecture} describes \tool{}'s architecture and \cref{sub:tool-implementation} gives an overview of its implementation.
More details about our implementation and the usage of \tool{} are given in \cref{app:implementation-and-usage}.

\subsection{Architecture}\label{sub:tool-architecture}
\begin{figure}
	\centering
	\includegraphics[width=\textwidth]{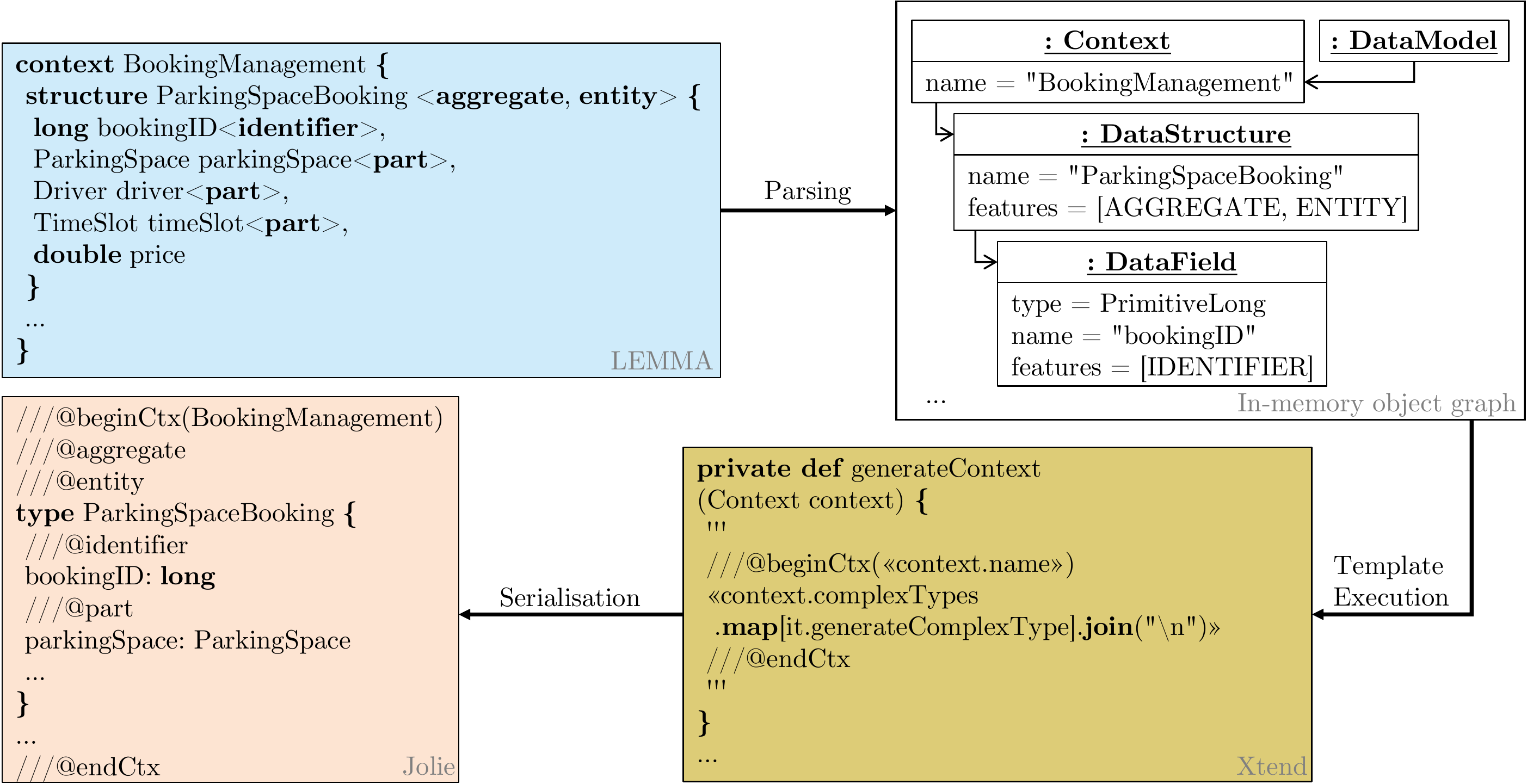}
	\caption{\tool{} phases to generate Jolie APIs from LEMMA domain models.}
	\label{fig:tool-phases}	
\end{figure}

As depicted in \cref{fig:tool-phases}, \tool{} consists of three phases to derive Jolie APIs from LEMMA domain models.

In the Parsing phase, \tool{} instantiates an in-memory object graph conforming to the metamodel of the DDML~\cite{Rademacher2020} from a given LEMMA domain model. The object graph allows systematic traversal of the model elements to map them to the corresponding Jolie code (cf. \cref{sec:encoding}) in the following Template Execution phase. As the phase name indicates, \tool{} relies on \emph{template-based code generation}~\cite{Combemale2017} to transform in-memory LEMMA domain models to Jolie. That is, we prescribe the target blocks of a Jolie program as strings involving static Jolie statements and dynamic variables which are evaluated at runtime to complement the prescribed target blocks with context-dependent information, e.g., the name of a bounded context in a specific LEMMA domain model. After template execution, the Serialisation phase stores the evaluated templates to physical files with valid Jolie code.

\subsection{Implementation Overview}\label{sub:tool-implementation}
We implemented \tool{} in Xtend\footnote{\url{https://www.eclipse.org/xtend}}, which is a Java dialect that integrates a sophisticated templating language (see below). Furthermore,  \tool{} relies on LEMMA's Java-based Model Processing Framework\footnote{\url{https://github.com/SeelabFhdo/lemma/tree/main/de.fhdo.lemma.model_processing}}, which aims to facilitate the development of model processors such as code generators. To this end, the framework provides built-in support for parsing models constructed with languages that are based on the Eclipse Modelling Framework~\cite{EMF_2008}---as is the case for all LEMMA modelling language including the DDML. Additionally, the framework prescribes a certain workflow for model processing and enables implementers to integrate with it using Java annotations.

Listing~\ref{lst:cgen-excerpt} describes the implementation of \tool{}'s code generation module which integrates with the Code Generation phase of LEMMA's Model Processing Framework. The module is responsible for template execution and the eventual serialisation of Jolie code (cf. \cref{sub:tool-architecture}).

\begin{listingXtend}{Xtend excerpt of \tool{}'s code generation module.}{lst:cgen-excerpt}{h}
§§@CodeGenerationModule(name="main")§§	#\label{lst:cgen-excerpt-anno}#
class GenerationModule extends AbstractCodeGenerationModule {#\label{lst:cgen-excerpt-class}#
	...
	override getLanguageNamespace() { return DataPackage.eNS_URI }	#\label{lst:cgen-excerpt-ns}#
	
	override execute(...) {			#\label{lst:cgen-excerpt-exec-begin}#
		val model = resource.contents.get(0) as DataModel	#\label{lst:cgen-excerpt-exec1}#
		val generatedContexts = model.contexts.map[it.generateContext]		#\label{lst:cgen-excerpt-exec2}#
		val baseFileName = FilenameUtils.getBaseName(modelFile)				#\label{lst:cgen-excerpt-exec3}#
		val targetFile = '''«targetFolder»«File.separator»«baseFileName».ol'''#\label{lst:cgen-excerpt-exec4}#
		return withCharset(#\##{targetFile -> generatedContexts.join("\n")},#\label{lst:cgen-excerpt-exec5}#
			StandardCharsets.UTF_8.name)								#\label{lst:cgen-excerpt-exec6}#
	}								#\label{lst:cgen-excerpt-exec-end}#
	
	private def generateContext(Context context) {'''	#\label{lst:template-excerpt-context-begin}#
		#///#@beginCtx(«context.name»)					#\label{lst:template-excerpt-beginctx}#
		«context.complexTypes.map[it.generateComplexType].join("\n")»#\label{lst:template-excerpt-delegate}#
		#///#@endCtx
	'''}											#\label{lst:template-excerpt-context-end}#

	private def dispatch generateComplexType(DataStructure structure) {'''
		«structure.generateType»			#\label{lst:template-excerpt-gentype}#
		«IF !structure.operations.empty»	#\label{lst:template-excerpt-geniface-begin}#
			«structure.generateInterface»		
		«ENDIF»								#\label{lst:template-excerpt-geniface-end}#
	'''}
}
\end{listingXtend}

A code generation module is a Java class with the \texttt{@Code\-Gen\-er\-a\-tion\-Mod\-ule} annotation that extends the \texttt{Ab\-stract\-Code\-Gen\-er\-a\-tion\-Mod\-ule} class (Lines~\ref{lst:cgen-excerpt-anno} and~\ref{lst:cgen-excerpt-class}). LEMMA's Model Processing Framework delegates to a code generation module after it parsed an input model in the modelling language supported by the module. To specify the supported language, a code generation module overrides the inherited \texttt{get\-Lan\-guage\-Namespace} method to return the language's namespace, which in the case of \tool{} is that of LEMMA's DDML (Line~\ref{lst:cgen-excerpt-ns}).

The entrypoint for code generation logic is the \texttt{ex\-ecute} method of a respective code generation module. It can access the in-memory object graph of a parsed model via the inherited \texttt{re\-source} attribute. Lines~\ref{lst:cgen-excerpt-exec-begin} to~\ref{lst:cgen-excerpt-exec-end} show the \texttt{ex\-ecute} method of \tool{}'s code generation module. In Line~\ref{lst:cgen-excerpt-exec1}, we retrieve the root of the model as an instance of the \texttt{Data\-Mod\-el} concept of the DDML's metamodel (cf. \cref{fig:tool-phases}). Next, we call the template method \texttt{gen\-er\-ate\-Con\-text} (see below) for each parsed \texttt{Con\-text} instance under the domain model root and gather the generated Jolie code as a list of strings in the \texttt{gen\-er\-ated\-Con\-texts} variable (Line~\ref{lst:cgen-excerpt-exec2}). In Lines~\ref{lst:cgen-excerpt-exec3} and~\ref{lst:cgen-excerpt-exec4}, we then determine the path of the generated Jolie file, which will be created in the given target folder and with the same base name as the input LEMMA domain model but with Jolie's extension ``ol''. Line~\ref{lst:cgen-excerpt-exec5} triggers the serialisation of the generated Jolie code via the inherited \texttt{with\-Char\-set} method.

Lines~\ref{lst:template-excerpt-context-begin} to~\ref{lst:template-excerpt-context-end} show the implementation of the template method \texttt{gen\-er\-ate\-Con\-text}. It expects an instance of the metamodel concept \texttt{Con\-text} as input (cf. \cref{fig:tool-phases}) and represents the starting point of each template execution since bounded contexts are the top-level elements in LEMMA domain models. An Xtend template is realized between a pair of three consecutive apostrophes within which it is whitespace-sensitive and preserves indentation. Within opening and closing guillemets, Xtend templates enable access to variables and computing operations, whose evaluation shall replace a certain template portion. Consequently, the expression \texttt{«con\-text.name»} in the template string in Line~\ref{lst:template-excerpt-beginctx} is at runtime replaced by the name of the bounded context passed to \texttt{gen\-er\-ate\-Con\-text}. For a bounded context with name ``Book\-ing\-Man\-age\-ment'', Line~\ref{lst:template-excerpt-beginctx} of the template will thus result in the generated Jolie code \texttt{///@be\-gin\-Ctx(Book\-ing\-Man\-age\-ment)} (cf. \cref{fig:tool-phases}).

To foster its overview and maintainability, we decomposed our template for Jolie APIs into several template methods following the specification of our encoding (cf. Sect.~\ref{sec:encoding}). As a result, the generation of Jolie code covering the internals of modelled bounded contexts happens in overloaded methods called \texttt{gen\-er\-ate\-Com\-plex\-Type}. Each of these methods derives Jolie code for a certain kind of LEMMA complex type, i.e., data structure, list, or enumeration. In Line~\ref{lst:template-excerpt-delegate}, the template delegates to the version of \texttt{gen\-er\-ate\-Com\-plex\-Type} for LEMMA data structures. Following our encoding, the method implements a template to map data structures to Jolie types (Line~\ref{lst:template-excerpt-gentype}) and interfaces in case the LEMMA data structure exhibits operation signatures (Lines~\ref{lst:template-excerpt-geniface-begin} to~\ref{lst:template-excerpt-geniface-end}).

The \tool{} source code is available on GitHub\footnote{\url{https://github.com/frademacher/lemma2jolie}}. In addition, we provide a publicly downloadable video illustrating \tool's practical capabilities\footnote{\url{https://bit.ly/3rTGysX}}.

\section{Related and Future Work}\label{sec:conclusion}

\paragraph{Related Work}

The maturity of MDE in research and practice as well as its ability to
effectively support the engineering of complex software systems~\cite{France2007}
has fostered the development of a variety of tools similar to
\tool{}~\cite{SCULPTOR,Kapferer2020,Kapferer2020b,Terzic2018,JDL2022}.
That is, they constitute code generators in the sense of MDE~\cite{Combemale2017}
and are capable to generate artefacts relevant to MSA engineering.
For this purpose, the tools process models constructed in a certain
modelling language.

However, and by contrast to \tool{}, the majority of related code
generators focuses on Java as target
technology~\cite{SCULPTOR,Terzic2018,JDL2022} and thus not on a programming
language specifically tailored to the challenges of microservice implementation. Reducing the semantic gap between the concepts of microservices and implementation languages is the reason for which new service-oriented languages like Ballerina and Jolie have been developed.
Furthermore, the modelling languages supported by related tools and hence
the generated code address only single concerns in MSA engineering, i.e.,
domain modelling~\cite{SCULPTOR,Kapferer2020} or the implementation and
provisioning of service APIs~\cite{Kapferer2020b,Terzic2018,JDL2022}. By contrast,
LEMMA's modelling languages offer an integrated solution to multi-concern modelling in MSA engineering,
by providing modelling languages dedicated to various viewpoints on microservice architectures (e.g., domain, service, and deployment)~\cite{Rademacher2020}.

\paragraph{Future Work}
The specified encoding (cf.~\cref{sec:encoding}) and its implementation
(cf.~\cref{sec:tool}) show the feasibility to integrate the LEMMA and Jolie
ecosystems. In future works we plan to extend this integration in several ways.

First, we plan to investigate the possibility of round-trip engineering (RTE), i.e.,
the bidirectional synchronisation of changes between LEMMA models and Jolie code.
This would enable, for example, domain experts and microservice developers to interact by using their views of interest (model vs implementation) but without risking that they fall out of sync.
While domain experts could continue to capture domain knowledge about a microservice
architecture in conceptual DDD domain models, developers could adapt
data types and APIs derived from those models using Jolie as their primary language.
Based on RTE, changes in Jolie code could then automatically be reflected in DDD domain
models and vice versa, with the option to immediately resolve potential conflicts in
domain understanding.

Second, we see potential for \tool{} to cover all phases in MSA
engineering, from domain-driven service design to implementation and
deployment. For example, we would like to extend \tool{} to deal also with the definition of access points (communication endpoints that define how APIs can be accessed), behaviours (implementations of services written in Jolie that accompany LEMMA models), and the generation of deployment configurations (e.g., configuration of infrastructural services, containerisation, and deployment plans for Kubernetes).
This potential is specifically fostered by both LEMMA and Jolie
constituting \emph{language-based approaches to MSA engineering}, which facilitates their integration. For example, we could extend LEMMA to include Jolie implementation code in service models.

\bibliographystyle{splncs04}
\bibliography{main}

\clearpage
\appendix

\section{Implementation and Usage of \toolname{} in Detail}
\label{app:implementation-and-usage}
Compared to \cref{sec:tool}, this appendix describes the implementation of \tool{} with LEMMA's Model Processing Framework in detail (cf. \cref{app:tool-implementation}). Furthermore, it explains the tool's usage (cf. \cref{app:tool-usage}).

\subsection{Implementation}\label{app:tool-implementation}
Like all LEMMA modelling languages, the DDML's implementation is based on the Eclipse Modelling Framework (EMF)~\cite{EMF_2008}. More precisely, the DDML's abstract syntax is implemented with Ecore, which is the metamodelling framework of EMF. The DDML's grammar for practical domain model construction was specified with the Xtext framework\footnote{\url{https://www.eclipse.org/Xtext}} for textual modelling languages. Xtext integrates with EMF and allows parser generation from grammar specifications. A generated Xtext parser is capable of translating textual model files to in-memory object graphs that constitute instantiations of the respective modelling language's metamodel. Such in-memory object graphs are then traversable for subsequent model processing steps using the EMF API.

LEMMA's Model Processing Framework makes the invocation of parsers for EMF-based modelling languages opaque to implementers, who instead gain direct access to in-memory object graphs as parsing results. The following paragraphs describe the implementation of each of \tool{}'s phases (cf. \cref{sub:tool-architecture}) with LEMMA's Model Processing Framework.

\subsubsection{Parsing}
In the Parsing phase, \tool{} parses a given LEMMA domain model into an in-memory object graph conforming to the metamodel of the DDML~\cite{Rademacher2020}. Using LEMMA's Model Processing Framework, we realised \tool{}'s programmatic entrypoint as shown in Listing~\ref{lst:tool-entrypoint}.

\begin{listingXtend}{Programmatic entrypoint of \tool{} written in Xtend.}{lst:tool-entrypoint}{h}
	class Lemma2Jolie extends AbstractModelProcessor {#\label{lst:tool-entrypoint-begin}#
		new() { super("lemma2jolie") }					#\label{lst:tool-entrypoint-constr}#
		def static void main(String[] args) { new Lemma2Jolie().run(args) }#\label{lst:tool-entrypoint-delegate}#
	}
\end{listingXtend}

LEMMA's Model Processing Framework supports the development of model processors as standalone executable Java applications based on the \texttt{Ab\-stract\-Mod\-el\-Pro\-cessor} class. Model processors extend this class (Line~\ref{lst:tool-entrypoint-begin}) and pass the name of a package to the framework in their constructor (Line~\ref{lst:tool-entrypoint-constr}). At runtime, the framework scans the passed package for annotated classes to invoke during model processor execution. Next to a constructor, entrypoints of model processors must implement a \texttt{main} method and delegate execution to the model processing framework by invoking the inherited \texttt{run} method with the given program arguments (Line~\ref{lst:tool-entrypoint-delegate}).

Next to an entrypoint, LEMMA model processors must implement a \emph{language description provider} to provide the model processing framework with information about the supported modelling language. Listing~\ref{lst:tool-langdescr} shows \tool{}'s language description provider.

\begin{listingXtend}{Xtend excerpt of \tool{}'s language description provider.}{lst:tool-langdescr}{h}
	§§@LanguageDescriptionProvider§§	#\label{lst:tool-langdescr-anno}#
	class LangDescriptionProvider implements LanguageDescriptionProviderI {#\label{lst:tool-langdescr-begin}#
		override getLanguageDescription(..., String namespaceOrExt) {#\label{lst:tool-langdescr-meth-begin}#
			return switch (namespaceOrExt) {
				case "data": new XtextLanguageDescription(DataPackage.eINSTANCE,	#\label{lst:tool-langdescr-value-check}#
					new DataDslStandaloneSetup)										#\label{lst:tool-langdescr-langdescr2}#
				...
			}
		}				#\label{lst:tool-langdescr-meth-end}#
	}
\end{listingXtend}

A language description provider is a class with the annotation \texttt{@Lan\-guage\-De\-scrip\-tion\-Pro\-vider}. Furthermore, the class must implement the interface \texttt{Lan\-guage\-De\-scrip\-tion\-Pro\-viderI} (Lines~\ref{lst:tool-langdescr-anno} and~\ref{lst:tool-langdescr-begin}) and override its \texttt{get\-Lan\-guage\-De\-scrip\-tion} method (Lines~\ref{lst:tool-langdescr-meth-begin} to~\ref{lst:tool-langdescr-meth-end}), which LEMMA's Model Processing Framework will invoke to retrieve information about a supported modelling language. Model processors like \tool{} can use the \texttt{namespace\-Or\-Ext} parameter to recognise the language of a given model file. Depending on the file format, the framework currently integrates language recognition based on XML namespaces or file extensions. Since our DDML is an Xtext-based modelling language and Xtext detects modelling languages from model files' extensions, \tool{} checks the \texttt{namespace\-Or\-Ext} parameter for the value ``data'' (Line~\ref{lst:tool-langdescr-value-check}), which is the file extension for LEMMA domain models.

Upon modelling language recognition, a model processor must return an instance of the \texttt{Lan\-guage\-De\-scrip\-tion} class. It informs the model processing framework about the parsing mechanism to use for an input model in a certain language. In Lines~\ref{lst:tool-langdescr-value-check} and~\ref{lst:tool-langdescr-langdescr2} of its language description provider, \tool{} returns a language description for Xtext-based modelling languages to the framework. Specifically, the corresponding \texttt{Xtext\-Lan\-guage\-De\-scrip\-tion} clusters an instance of the \texttt{Data\-Dsl\-Stan\-dalone\-Setup} class generated by Xtext. From this class, the model processing framework is able to trigger and control the parsing of LEMMA domain models into DDML-conform in-memory object graphs (cf. \cref{sub:tool-architecture}).

\subsubsection{Template Execution}
In this phase, \tool{} executes a template for Jolie APIs on the in-memory object graph of a parsed LEMMA domain model. We used the integrated templating language of Xtend to formulate the template. Listing~\ref{lst:template-excerpt} shows an excerpt of the template.

\begin{listingXtend}{Excerpt of our template for Jolie APIs in Xtend's templating language.}{lst:template-excerpt}{h}
	private def generateContext(Context context) {'''	#\label{app:lst:template-excerpt-context-begin}#
		#///#@beginCtx(«context.name»)					#\label{app:lst:template-excerpt-beginctx}#
		«context.complexTypes.map[it.generateComplexType].join("\n")»#\label{app:lst:template-excerpt-delegate}#
		#///#@endCtx
	'''}											#\label{app:lst:template-excerpt-context-end}#
	
	private def dispatch generateComplexType(DataStructure structure) {'''
		«structure.generateType»			#\label{app:lst:template-excerpt-gentype}#
		«IF !structure.operations.empty»	#\label{app:lst:template-excerpt-geniface-begin}#
			«structure.generateInterface»		
		«ENDIF»								#\label{app:lst:template-excerpt-geniface-end}#
	'''}
\end{listingXtend}

Lines~\ref{app:lst:template-excerpt-context-begin} to~\ref{app:lst:template-excerpt-context-end} show the implementation of the template method \texttt{gen\-er\-ate\-Con\-text}. It expects an instance of the metamodel concept \texttt{Con\-text} as input (cf. \cref{fig:tool-phases}) and represents the starting point of each template execution since bounded contexts are the top-level elements in LEMMA domain models. An Xtend template is realized between a pair of three consecutive apostrophes within which it is whitespace-sensitive and preserves indentation. Within opening and closing guillemets, Xtend templates enable access to variables and computing operations, whose evaluation shall replace a certain template portion. Consequently, the expression \texttt{«con\-text.name»} in the template string in Line~\ref{app:lst:template-excerpt-beginctx} is at runtime replaced by the name of the bounded context passed to \texttt{gen\-er\-ate\-Con\-text}. For a bounded context with name ``Book\-ing\-Man\-age\-ment'', Line~\ref{app:lst:template-excerpt-beginctx} of the template will thus result in the generated Jolie code \texttt{///@be\-gin\-Ctx(Book\-ing\-Man\-age\-ment)} (cf. \cref{fig:tool-phases}).

To foster its overview and maintainability, we decomposed our template for Jolie APIs into several template methods following the specification of our encoding (cf. Sect.~\ref{sec:encoding}). As a result, the generation of Jolie code covering the internals of modelled bounded contexts happens in overloaded methods called \texttt{gen\-er\-ate\-Com\-plex\-Type}. Each of these methods derives Jolie code for a certain kind of LEMMA complex type, i.e., data structure, list, or enumeration. In Line~\ref{app:lst:template-excerpt-delegate}, the template in Listing~\ref{lst:template-excerpt} delegates to the version of \texttt{gen\-er\-ate\-Com\-plex\-Type} for LEMMA data structures. Following our encoding, the method implements a template to map data structures to Jolie types (Line~\ref{app:lst:template-excerpt-gentype}) and interfaces in case the LEMMA data structure exhibits operation signatures (Lines~\ref{app:lst:template-excerpt-geniface-begin} to~\ref{app:lst:template-excerpt-geniface-end}).

\subsubsection{Serialisation}
In its last phase, \tool{} serialises the results from template execution to physical files with Jolie code. To this end, we leveraged LEMMA's Model Processing Framework to implement a code generation module. Listing~\ref{app:lst:cgen-excerpt} shows an excerpt of its Xtend implementation.

\begin{listingXtend}{Xtend excerpt of \tool{}'s code generation module.}{app:lst:cgen-excerpt}{h}
	§§@CodeGenerationModule(name="main")§§	#\label{app:lst:cgen-excerpt-anno}#
	class GenerationModule extends AbstractCodeGenerationModule {#\label{app:lst:cgen-excerpt-class}#
		...
		override getLanguageNamespace() { return DataPackage.eNS_URI }	#\label{app:lst:cgen-excerpt-ns}#

		override execute(...) {			#\label{app:lst:cgen-excerpt-exec-begin}#
			val model = resource.contents.get(0) as DataModel	#\label{app:lst:cgen-excerpt-exec1}#
			val generatedContexts = model.contexts.map[it.generateContext] #\label{app:lst:cgen-excerpt-exec2}#
			val baseFileName = FilenameUtils.getBaseName(modelFile)		#\label{app:lst:cgen-excerpt-exec3}#
			val targetFile = '''«targetFolder»«File.separator»«baseFileName».ol'''#\label{app:lst:cgen-excerpt-exec4}#
			return withCharset(#\##{targetFile -> generatedContexts.join("\n")},#\label{app:lst:cgen-excerpt-exec5}#
				StandardCharsets.UTF_8.name)								#\label{app:lst:cgen-excerpt-exec6}#
		} #\label{app:lst:cgen-excerpt-exec-end}#
		
		#/* cf. Listing~\ref{lst:template-excerpt} */#
		private def generateContext(Context context) { ... }
		private def dispatch generateComplexType(DataStructure structure) { ... }
	}
\end{listingXtend}

A code generation module in the sense of LEMMA's Model Processing framework is a Java class with the \texttt{@Code\-Gen\-er\-a\-tion\-Mod\-ule} annotation and extending the \texttt{Ab\-stract\-Code\-Gen\-er\-a\-tion\-Mod\-ule} class (Lines~\ref{app:lst:cgen-excerpt-anno} and~\ref{app:lst:cgen-excerpt-class}). The model processing framework delegates to a code generation module after it parsed an input model with the namespace of the modelling language supported by the module. As \tool{} parses LEMMA domain models (cf. \cref{fig:tool-phases}), the code generation module returns the namespace of LEMMA's DDML to the framework (Line~\ref{app:lst:cgen-excerpt-ns}).

The entrypoint for code generation logic is the \texttt{ex\-ecute} method of a respective code generation module. The in-memory object graph of the parsed model is accessible via the inherited \texttt{re\-source} attribute. Lines~\ref{app:lst:cgen-excerpt-exec-begin} to~\ref{app:lst:cgen-excerpt-exec-end} show the \texttt{ex\-ecute} method in the code generation module of \tool. In Line~\ref{app:lst:cgen-excerpt-exec1}, we retrieve the root of the model as an instance of the \texttt{Data\-Mod\-el} concept of the DDML's metamodel (cf. \cref{fig:tool-phases}). Next, we call the template method \texttt{gen\-er\-ate\-Con\-text} (cf. Listing~\ref{lst:template-excerpt}) for each parsed \texttt{Con\-text} instance under the domain model root and gather the generated Jolie code as a list of strings in the \texttt{gen\-er\-ated\-Con\-texts} variable (Line~\ref{app:lst:cgen-excerpt-exec2}).

Finally, we determine the path of the file for the generated Jolie code, which will be created in the given target folder and with the same base name as the input LEMMA domain model but with the extension ``ol''  (Lines~\ref{app:lst:cgen-excerpt-exec3} and~\ref{app:lst:cgen-excerpt-exec4}). The serialisation of the generated Jolie code is triggered by invoking the inherited \texttt{with\-Char\-set} method and returning its results to the framework. The method expects a map of file paths and contents, and the target encoding as argument. For \tool{}'s code generation module, the first argument associates the previously assembled path of the file for the generated Jolie code with the generated code concatenated in a string separated by line breaks (Line~\ref{app:lst:cgen-excerpt-exec5}). As the second argument of \texttt{with\-Char\-set}, we pass an identifier for UTF-8 encoding (Line~\ref{app:lst:cgen-excerpt-exec6}).

\subsection{Usage}\label{app:tool-usage}
\tool{} integrates a commandline interface which is executable with Java 11 or greater. \tool{} can be compiled to a standalone Java archive from its GitHub sources and run on physical hardware. Alternatively, it is possible to execute \tool{} in a Docker container\footnote{\url{https://www.docker.com}} for which we provide a dedicated Dockerfile on GitHub. In either case, the commandline invocation follows the pattern shown in Listing~\ref{lst:cli-pattern}.

\begin{listingXtend}{Commandline pattern for invoking \tool.}{lst:cli-pattern}{h}
	java -jar lemma2jolie.jar -s <LEMMA_MODEL> -t <JOLIE_FOLDER>
\end{listingXtend}

The commandline option \texttt{-s} expects the path of a LEMMA domain model. Hence, \tool{} assumes upfront construction of a LEMMA domain model, e.g., with LEMMA's editor plugins for the Eclipse IDE which provide sophisticated modelling support including syntax highlighting, code completion and cross-referencing~\cite{Rademacher2020}. The commandline option \texttt{-t} then points to the target folder for the generated Jolie file (cf. Sect.~\ref{sub:tool-implementation})
\end{document}